\documentclass[preprint2]{aastex}
\input{english.sty}
\def\etal{{\it et\thinspace al.\/~}}

\def\beq#1{\begin{equation}\label{#1}}
\def\eeq{\end{equation}}
\def\beqa#1{\begin{eqnarray}\label{#1}}
\def\eeqa{\end{eqnarray}}

\def\tento#1{\times 10^{#1}}

\def\Ms{\ M_{\odot}}     

\def\K{{\rm \ K}}

\def\cm{{\rm \ cm}}

\def\pc{{\rm \ pc}}

\def\yrs{{\rm \ years}}

\def\HH{H$_2$ }
\def\H2p{H$_2^+$ }
\def\Hp{H$^+$ }
\def\Hm{H$^-$ }
\def\Hep{He$^+$ }
\def\Hepp{He$^{++}$ }

\def\HH{H$_2$ }
\def\H2p{H$_2^+$ }
\def\HHp{H$_2^+$ }
\def\Hp{H$^+$ }
\def\Hm{H$^-$ }
\def\Hep{He$^+$ }
\def\Hepp{He$^{++}$ }

\def\mH2p{H_2^+}

\def\gtsima{$\; \buildrel > \over \sim \;$}
\def\ltsima{$\; \buildrel < \over \sim \;$}
\def\prosima{$\; \buildrel \propto \over \sim \;$}
\def\gsim{\lower.7ex\hbox{\gtsima}}
\def\lsim{\lower.7ex\hbox{\ltsima}}
\def\simgt{\lower.7ex\hbox{\gtsima}}
\def\simlt{\lower.7ex\hbox{\ltsima}}
\def\simpr{\lower.7ex\hbox{\prosima}}


\begin{document}
\title{The Formation and Fragmentation of Primordial Molecular Clouds}
\author{Tom Abel$^{1}$, Greg L. Bryan$^{2,3}$ and Michael L. Norman$^{4,5}$\\
{\it $^1$Harvard Smithsonian Center for Astrophysics, MA, US--02138 Cambridge}\\
{\it $^2$Massachusetts Institute of Technology, MA, US--02139 Cambridge}\\
{\it $^3$Hubble Fellow}\\
{\it $^4$LCA, NCSA, University of Illinois, US--61801 Urbana/Champaign}\\
{\it $^5$Astronomy Department, University of Illinois, Urbana/Champaign} }
\begin{abstract} Many questions in physical cosmology regarding the thermal
  history of the intergalactic medium, chemical enrichment,
  reionization, etc.  are thought to be intimately related to the
  nature and evolution of pregalactic structure.  In particular the
  efficiency of primordial star formation and the primordial IMF are
  of special interest.  We present results from high resolution
  three--dimensional adaptive mesh refinement simulations that follow
  the collapse of primordial molecular clouds and their subsequent
  fragmentation within a cosmologically representative volume.
  Comoving scales from 128 kpc down to 1 pc are followed accurately.
  Dark matter dynamics, hydrodynamics and all relevant chemical and
  radiative processes (cooling) are followed self-consistently for a
  cluster normalized CDM structure formation model.  Primordial
  molecular clouds with $\sim 10^5$ solar masses are assembled by
  mergers of multiple objects that have formed hydrogen molecules in
  the gas phase with a fractional abundance of $\lsim 10^{-4}$.  As
  the subclumps merge cooling lowers the temperature to $\sim200$ K
  in a ``cold pocket'' at the center of the halo. Within this cold pocket, a
  quasi--hydrostatically contracting core with mass $\sim 200\Ms$ and
  number densities $\gsim 10^5 \cm^{-3}$ is found.  We find that less
  than 1\% of the primordial gas in such small scale structures cools
  and collapses to sufficiently high densities to be available for
  primordial star formation.  
  Furthermore, it is worthwhile to note that this study achieved the
  highest dynamic range covered by structured adaptive mesh techniques
  in cosmological hydrodynamics to date. \vspace{0.3cm}
\end{abstract}

\thispagestyle{empty}
\section{Introduction}

Saslaw and Zipoy (1967) realized the importance of gas phase \HH
molecule formation in primordial gas for the formation of
proto--galactic objects.  Employing this mechanism in Jeans unstable
clouds, Peebles and Dicke (1968) formulated their model for the
formation of primordial globular clusters. Further pioneering studies
in this subject were carried out by Takeda \etal (1969), Matsuda \etal
(1969), and Hirasawa \etal (1969) who followed in detail the gas
kinetics in collapsing objects and studied the possible formation of
very massive objects (VMO's).  In the 1980's the possible cosmological
consequences of population III star formation were assessed (Rees and
Kashlinsky 1983; Carr \etal 1984; Couchman and Rees 1986). In
particular Couchman and Rees (1986) discussed first structure
formation within the standard cold dark matter model.  Their
main conclusions were that the first objects might reheat and reionize
the universe, raise the Jeans mass and thereby influence subsequent
structure formation.

Early studies focused on the chemical evolution and cooling of
primordial clouds by solving a chemical reaction network within highly
idealized collapse models (cf.  Hirasawa 1969; Hutchins 1976; Palla
\etal 1983; MacLow and Shull 1986; Puy \etal 1996; Tegmark \etal
1997).  Some hydrodynamic aspects of the problem were studied in
spherical symmetry by Bodenheimer (1986) and Haiman, Thoul and Loeb
(1996).  Recently multi--dimensional studies of first structure
formation have become computationally feasible (Abel 1995; Anninos \&
Norman 1996; Zhang \etal1997; Gnedin
\& Ostriker 1997; Abel \etal 1998a,1998b; Bromm \etal~1999). These
investigations have provided new insights into the inherently
multidimensional, nonlinear, nonequilibrium physics which determine
the collapse and fragmentation of gravitationally and thermally
unstable primordial gas clouds.

In Abel, Anninos, Norman \& Zhang 1998a (hereafter AANZ) we presented
the first self-consistent 3D cosmological hydrodynamical simulations
of first structure formation in a standard cold dark matter--dominated
(SCDM) universe. These simulations included a careful treatment of the
formation and destruction of \HH---the dominant coolant in low mass
halos ($M_{tot}=10^5-10^8 M_{\odot}$) which collapse at high redshifts
($z \sim 30-50$).  Among the principal findings of that study were:
(1) appreciable cooling only occurs in the cores of the high density
spherical knots located at the intersection of filaments; (2) good
agreement was found with semi-analytic predictions (Abel 1995; Tegmark
\etal 1997) of the minimum halo mass able to cool and collapse to
higher densities; (3) only a small fraction ($< 10\%$) of the bound
baryons are able to cool promptly, implying that primordial Pop III
star clusters may have very low mass. Due to the limited spatial
resolution of the those simulations ($\sim 1$ kpc {\em comoving}), we
were unable to study the collapse to stellar densities and address the
nature of the first objects formed.

In this paper we present new, higher- resolution results using the
powerful numerical technique of adaptive mesh refinement (AMR, Bryan
\& Norman 1997; Norman \& Bryan 1999) which has shed some light on how
the cooling gas fragments.  With an effective dynamic range of
$262,144$ the numerical simulations presented here are the highest
resolution simulations in cosmological hydrodynamics to date. Although
we are not yet able to form individual protostars, we are able to
resolve the collapsing protostellar cloud cores which must inevitably
form them. We find the cores have typical masses $\sim 200 M_{\odot}$,
sizes $\sim 0.3$ pc, and number densities $n \geq 10^5$
cm$^{-3}$--similar to dense molecular cloud cores in the Milky Way
with one vital difference: the molecular hydrodgen fraction is $\sim 5
\times 10^{-4}$, meaning the cores evolves very differently from
Galactic cores.

The plan of this paper is as follows. The simulations are briefly
described in Sec. 2. Results are presented in Sec. 3. The properties
and fate of the primordial protostellar cloud are discussed in
Sec. 4. Conclusions follow in Sec. 5.  Results of a broader survey of
simulations will be reported in Abel, Bryan \& Norman (1999).

\section{Simulations}

The three dimensional adaptive mesh refinement calculations presented
here use for the hydrodynamic portion an algorithm very similar to the
one described by Berger and Collela (1989).  The code utilizes an
adaptive hierarchy of grid patches at various levels of resolution.
Each rectangular grid patch covers some region of space in its parent
grid needing higher resolution, and may itself become the parent grid
to an even higher resolution child grid. Our general implementation of
AMR places no restriction on the number of grids at a given level of
refinement, or the number of levels of refinement.  However, we do
restrict the refinement factor -- the ratio of parent to child mesh
spacing -- to be an integer (chosen to be 2 in this work).  The dark
matter is followed with methods similar to the ones presented by
Couchman (1991). Furthermore, the algorithm of Anninos \etal (1997) is
used to solve the time--dependent chemistry and cooling equations for
primordial gas given in Abel \etal (1997).  More detailed descriptions
of the code are given in Bryan \& Norman (1997, 1999), and Norman \&
Bryan (1999).

The simulations are initialized at redshift 100 with density
perturbations of a SCDM model with $\Omega_B = 0.06$, $h=0.5$, and
$\sigma_8=0.7$. The abundances of the 9 chemical species (H, \Hp, \Hm,
He, \Hep, \Hepp, \HH, \HHp, e$^-$) and the temperature are initialized
as discussed in Anninos and Norman (1996). After a collapsing
high--$\sigma$ peaks has been identified in a low resolution run, the
simulation is reinitialized with multiple refinement levels covering
the Langrangian volume of the collapsing structure.  The mass
resolution in the initial conditions within this region are $0.53
(8.96) \Ms$ in the gas (dark matter). The refinement criteria ensure
that: (1) the local Jeans length is resolved by at least 4 grid zones,
and (2) that no cell contains more than 4 times the initial mass
element ($0.53\Ms$). We limit the refinement to 12 levels within a
$64^3$ top grid which translates to a maximum dynamic range of
$64\times 2^{12}=262,144$.

\section{Results}

We find that primordial molecular clouds are only formed at the
intersection of filaments, in agreement with the results of AANZ.
The evolution of these primordial molecular clouds is marked by
frequent mergers yielding highly complex velocity and density fields
within the ``virial'' radius.  In the following three sections, we
first describe the evolution of these objects and then their morphology
and structure.

\subsection{Formation of the First Objects}

\begin{figure*}\vspace{0.4cm}
\epsscale{2}
\plotone{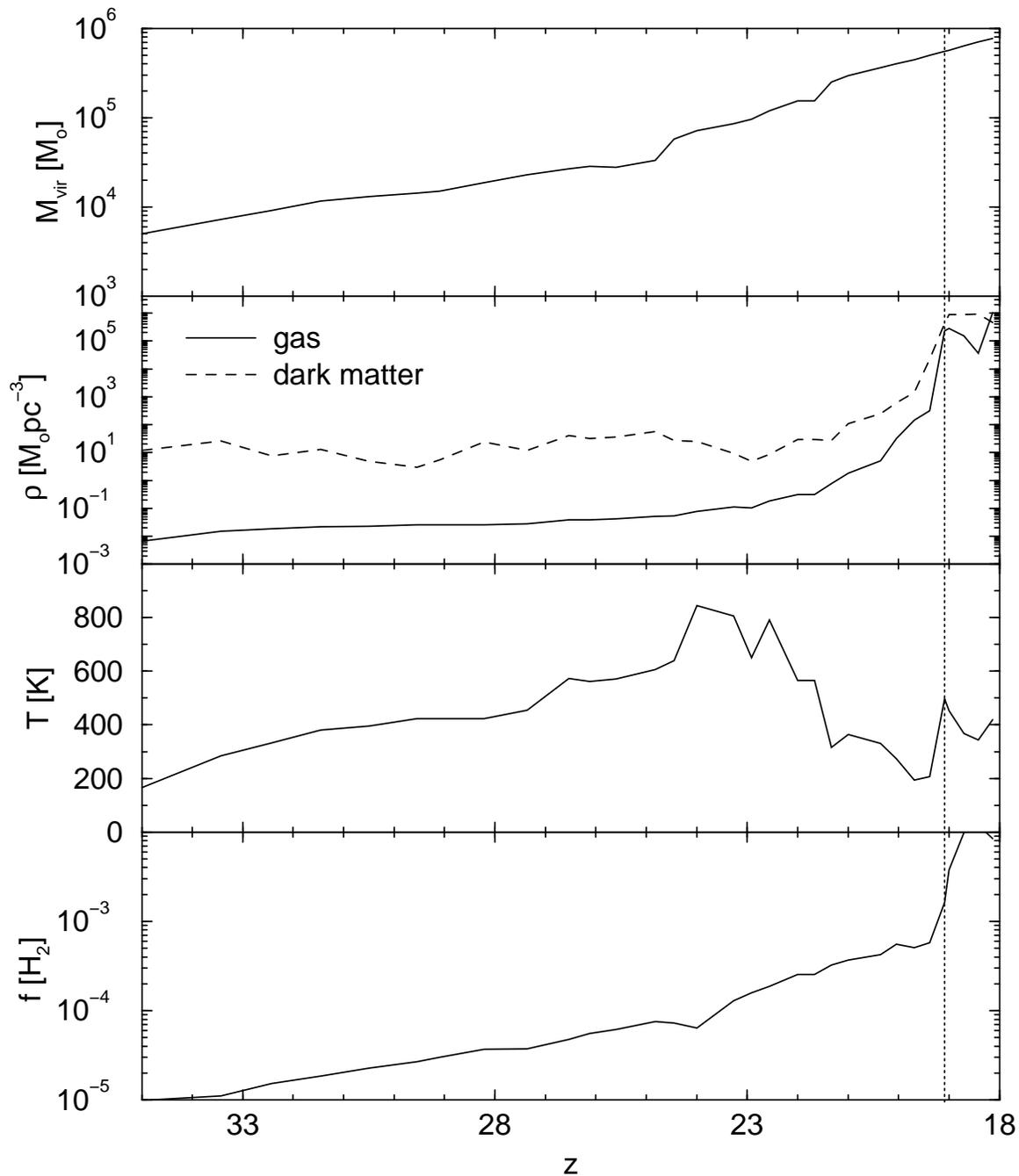}\vspace{-0.4cm}
\caption{ The top panel shows the evolution of the virial
mass of the most massive clump as a function of redshift.  The remaining
panels show the density (both dark and baryonic), the temperature, and
the molecular hydrogen mass fraction at the central point of that clump.  The
central point is defined as the point with the highest baryon density.
Clearly the finite gas pressure prevents baryons from clumping as
much as the dark matter at redshifts $\gsim 23$.  The vertical line at $z=19.1$
indicates where our numerical model breaks down.} \vspace{.2cm}
\label{evolution}
\end{figure*}

To illustrate the physical mechanisms at work during the formation of
the first cosmological object in our simulation, we show the evolution
of various quantities in Figure~\ref{evolution}.  The top panel of
this plot shows the virial mass of the largest object in the
simulation volume.  We divide the evolution up into four intervals.
In the first, before a redshift of about 35, the Jeans mass in the
baryonic component is larger than the mass of any non-linear
perturbation.  Therefore, the only collapsed objects are dark-matter
dominanted, and the baryonic field is quite smooth.  (We remind the
reader that a change in the adopted cosmological model would modify
the timing, but not the nature, of the collapse.)

In the second epoch, $23 < z < 35$, as the non-linear mass increases, the
first baryonic objects collapse.  However, these cannot efficiently
cool and the primordial entropy of the gas prevents dense cores from
forming.  This is shown in the second frame of figure~\ref{evolution}
by a large gap between the central baryonic and dark matter densities
(note that while the dark matter density is limited by resolution, the
baryonic is not, so the true difference is even larger).  As mergers
continue and the mass of the largest clump increases, its temperature
also grows, as shown in the third panel of this figure.  The \HH
fraction also increases (bottom panel).

By $z \sim 23$, enough \HH has formed (a few $\times 10^{-4}$), and the
temperature has grown sufficiently high that cooling begins to be
important.  During this third phase, the central temperature decreases
and the gas density increases.  However, the collapse is somewhat
protracted because around this point in the evolution, the central
density reaches $n \sim 10^4$ cm$^{-2}$, and the excited states of \HH
are in LTE.  This results in a cooling time which is nearly
independent of density rather than in the low--density limit where
$t_{cool} \sim \rho^{-1}$ (e.g. Lepp \& Shull 1983).

Finally, at $z \sim 19$, a very small dense core forms and reaches the
highest resolution that we allowed the code to produce.  It is
important to note that at this point, the maximum gas density in the
simulations exceeds $10^8\cm^{-3}$, and at these densities, 3--body
formation of molecular hydrogen will become dominant (see Palla
\etal~1983). Also, the assumption of optical thin cooling begins to
break down and radiative transfer effects become important.
Therefore, only simulation results at and above this redshift will be
discussed. It is worthwhile to note that the simulations presented
here are physics rather than resolution limited.

\subsection{Morphology}

The increase in dynamic range by $\sim 1000$ in the simulations
presented here as compared to AANZ allow us to investigate the
details of the fragmentation process in detail.  Visualizations of the
gas density and temperature on two different scales at $z=19.1$ are
shown in the color plate~\ref{color_p}.  In the upper left panel the
velocity field is shown superimposed on the density. The
$5\tento{5}\Ms$ structure forms at the intersection of two filaments
with overdensities of $\sim 10$.  Most of the mass accretion occurs
along these filaments.  The complexity of the velocity field is
evident; the accretion shock is highly aspherical and of varying
strength.  Within the virial radius ($r = 106$ pc), there are a
number of other cooling regions.
The right-hand panels zooms in on the collapsing fragment.
(note that the smallest resolution element ($0.02\pc$) in the
simulations is still 1600 times smaller than the slice shown in the
right panels).  The small fragment in the center of this image has a
typical overdensity of $\gsim 10^{6}$ and a mass of $\sim 200\Ms$.

\subsection{Profiles}

\begin{figure*}
\vspace{0.3cm}
\epsscale{1.5}
\plotone{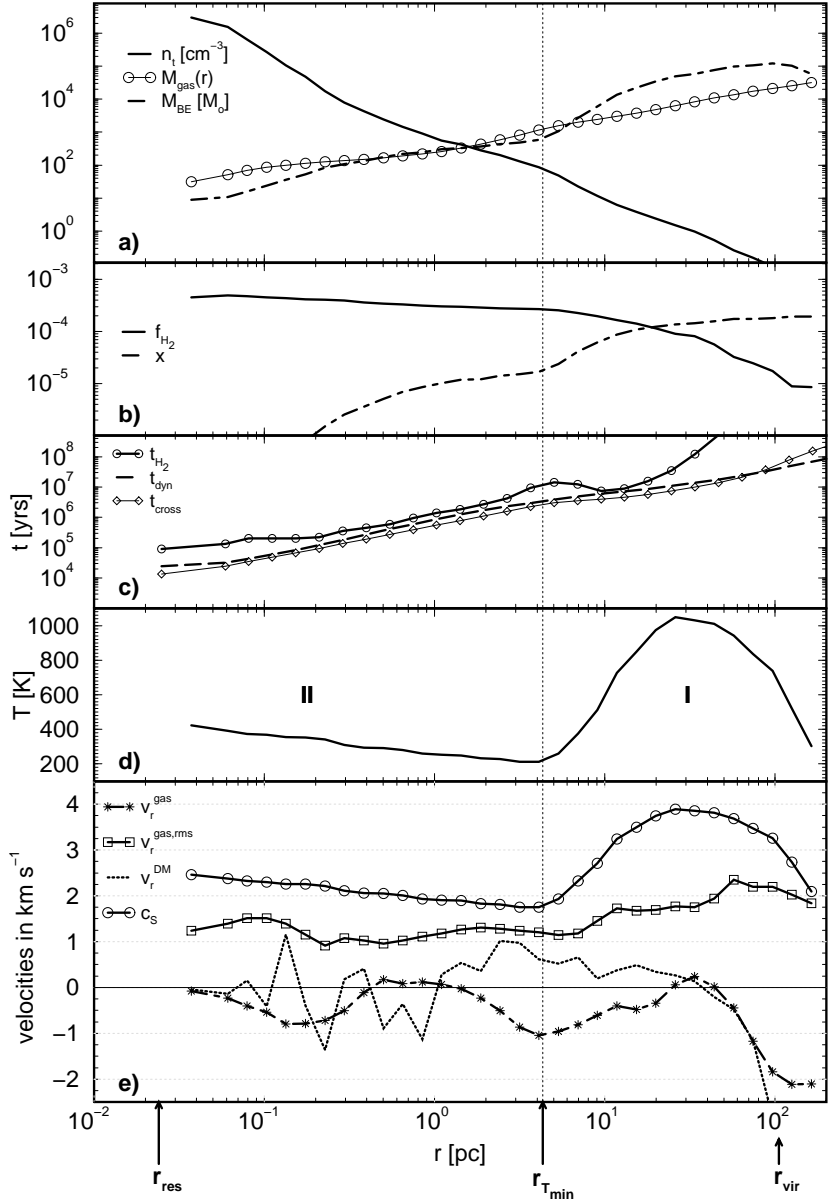}\vspace{+0.2cm}\label{profile}
\caption{ Spherically averaged mass weighted profiles around the
  baryon density peak shortly before a well defined fragment forms
  (z=19.1). Panel a) shows the baryonic number density, enclosed gas
  mass in solar mass, and the local Bonnor--Ebert mass ($\approx
  27\Ms T_K^{1.5}/\sqrt{n}$).  Panel b) plots the molecular hydrogen
  number fraction $f_{H_2}$ and the free electron number fraction $x$.
  The \HH cooling time $t_{H_2}$, the time it takes a sound wave to
  travel to the center, $t_{cross}$, and the free--fall time
  $t_{ff}=[3\pi/(32G\rho)]^{1/2}$ are given in panel c). Panel d)
  gives the temperature in Kelvin as a function of radius.  The bottom
  panel gives the local sound speed, $c_s$ (solid line with circles),
  the rms radial velocities of the dark matter (dashed line) and the
  gas (dashed line with asterisks) as well as the rms gas velocity
  (solid line with square symbols).
  The vertical dotted line indicates the radius at which the gas has
  reached its minimum temperature allowed by \HH cooling ($\sim 5pc$).
  The virial radius of the $5.6\tento{6}\Ms$ halo is $106\pc$. The
  cell size on the finest grid corresponds to $0.024\pc$. Note that
  the simulation box size corresponds to 6.4 proper kpc.}
\end{figure*}

\begin{figure*}
\plotone{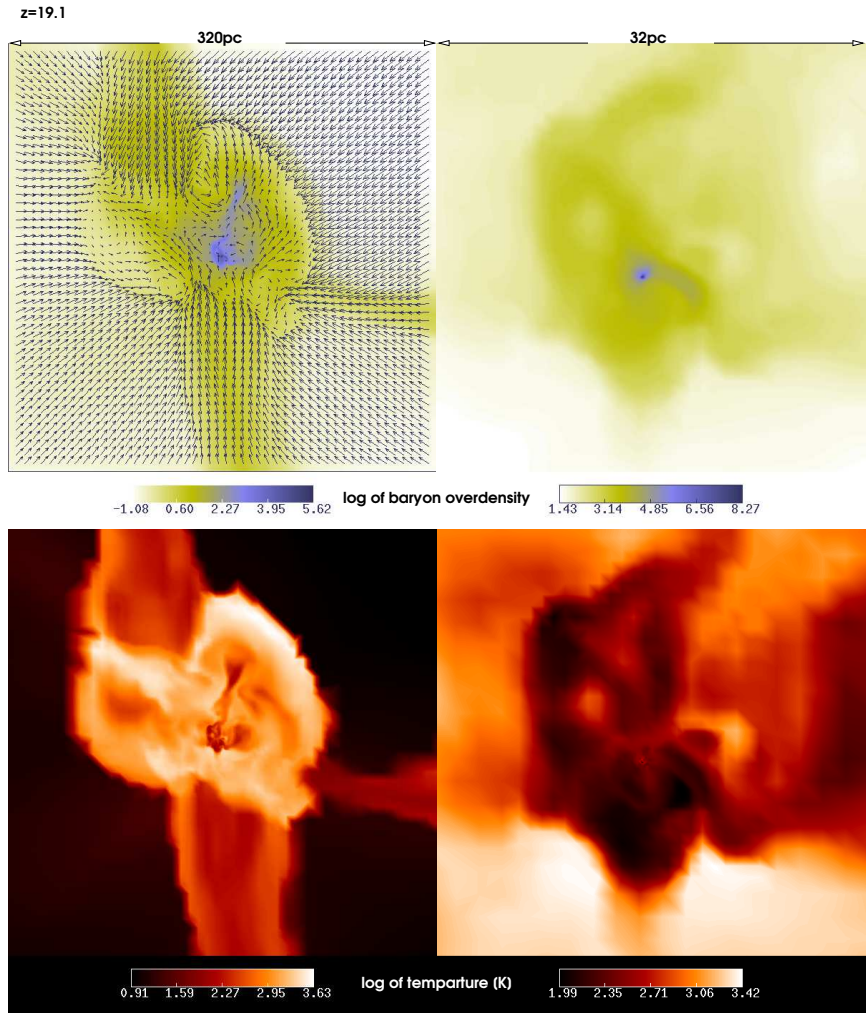}
\caption{Gas density and temperature in the first cosmological objects
  expected to form in hierarchical structure formation scenarios.  The
  upper panels show the log of the baryonic overdensity in a slice
  through the point of highest gas density at a scale of $320\pc$
  (left) and $32\pc$ (right). The lower panels give the corresponding
  plots of the log of the gas temperature. Additionally the velocity
  field is also visualized in the upper left panel. Note that the
  computational volume simulated is 20$^3$ times larger than the left
panels. }\label{color_p} 
\end{figure*}

Despite the complex structure of the primordial molecular clouds much
of their structure can be understood from spherical profiles of the
physical quantities, particularly for the dense central core which is
nearly spherical.  Figure~\ref{profile} shows mass-weighted, spherical
averages of various quantities around the densest cell found in the
simulation at redshift 19.1. Panel a) plots the baryon number density,
enclosed baryon mass, and local Bonnor-Ebert mass\footnote{{\sl
Bonnor-Ebert mass} is the analog of the Jeans Mass but assuming an
isothermal ($\rho\propto r^{-2}$) instead of a uniform density
distribution.}  $\approx 27\Ms T_K^{1.5}/\sqrt{n}$ versus radius.
Panel b) plots the abundances of \HH and free electrons.  Panel c)
compares three timescales defined locally: the \HH cooling time
$t_{H_2}$, the freefall time $t_{ff}=[3\pi/(32G\rho)] ^{1/2}$, and the
sound crossing time $t_{cross}=r/c_s=7.6\tento{6} r_{pc}/\sqrt{T_K}$
yrs.  In panel d) we identify two distinct regions---labeled I and
II---as defined by the temperture profile.  Region I ranges from
outside the virial radius to $r_{T_{min}}\sim 5\pc$, the radius at
which the infalling material has cooled down to $T_{min} \sim
200\K$--near the minimum temperature allowed by \HH cooling. Within
region I, the temperature profile reflects, in order of decreasing
radius, cosmic infall, shock virialization, adiabatic heating in a
settling zone, and an \HH cooling flow.  In region II, the temperature
slowly rises from $T_{min}$ to $\sim 400 K$ due to adiabatic heating.

For most of region I the \HH cooling time $t_{H_2}$ is comparable to
the free--fall time, as is illustrated in panel c) of
Figure~\ref{profile}. The \HH number fraction rises from $7\tento{-6}$
to $2\tento{-4}$ as the free electron fraction drops from
$2\tento{-4}$ to $2\tento{-5}$.
At $r_{T_{min}}$, the sound crossing time becomes substantially
shorter than the cooling time. This suggests that region II is
contracting quasi--hydrostatically on the cooling time scale, which
approaches its constant high--density value at small radii.
This constant cooling time of $\sim 10^5\yrs$ sets the time scale of
the evolution of the fragment until it can turn fully molecular via
three body associations.
Inside $r \sim 0.3$ pc, the enclosed baryonic mass of $\sim 200
M_{\odot}$ exceeds the local Bonnor-Ebert mass, implying this material
is gravitationally unstable.  However, due to the inefficient cooling,
its collapse is subsonic (panel e).  The radius where $M > M_{BE}$
defines our protostellar cloud core.

\section{Discussion}

Many interesting features of the collapsing and fragmenting
``primordial molecular cloud'' are identified. Most notable is the
formation of an initially quasi--hydrostatically contracting core of
$\sim 200\Ms$ which becomes gravitationally unstable.  We argue that
this is a characteristic mass scale for core formation mediated by \HH
cooling. Substituting into the formula for the Bonnor-Ebert mass
$T_{min}$ and $n_{LTE}$ we get $240 M_{\odot}$.
 
What will be the fate of the collapsing core? Within the core the
number densities increase from $10^{5}$ to $10^8\cm^{-3}$. For
densities $\gsim 10^8\cm^{-3}$, however, three--body formation of \HH
will become the dominant formation mechanism, transforming all
hydrogen into its molecular form (Palla \etal1983).  Our chemical
reaction network does not include this reaction and the solution
cannot be correct at $r\lsim 0.1\pc$. The most interesting effect of
the three--body reaction is that it will increase the cooling rate by
a factor $\sim 10^3$, leading to a further dramatic density
enhancement within the core. This will decrease the dynamical
timescales to $\ll 100\yrs$, effectively decoupling the evolution of
the fragment from the evolution of its host primordial molecular
cloud. Therefore, it is a firm conclusion that only the gas within
these cores can participate in population III star formation.

Omukai \& Nishi (1998) have simulated the evolution of a collapsing,
spherically symmetric primordial cloud to stellar density including
all relevant physical processes. Coincidentally, their initial
conditions are very close to our final state. Based on their results,
we can say that if the cloud does not break up, a massive star will be
formed.  Adding a small amount of angular momentum to the core does
not change this conclusion (Bate 1998). A third possibility is that
the cloud breaks up into low mass stars via thermal instability in the
quasi-hydrostatic phase. Silk (1983) has argued that, due to the
enhanced cooling from the 3--body produced \HH, fragmentation of this
core might continue until individual fragments are opacity limited
(i.e. they become opaque to their cooling radiation). Exploring which
of these scenarios is correct will have to await yet higher resolution
simulations including the effects of radiative transfer.  It will also
be interesting to examine the possible effects of molecular HD which,
although much less abundant, is a much more efficient coolant at low
temperatures.

How many cores are formed in our halo? Because our timestep contracts
rapidly once the first core forms, we are not yet able to answer this
question definitively. An earlier, less well resolved simulation
yielded $5-6$ cores by $z=16.5$, suggesting that multiple cores do
form. We speculate that the total number of cores to eventually form
will be proportional to the total amount of cooled gas. However, the
first star in a given halo will most likely always be formed close to
its center where the dynamical timescale is shortest. The cooling
timescale at $r_{T_{min}}$ in Figure 2. of $\gsim 10^6 \yrs$ should
roughly correspond to the typical formation time of fragments.  During
this time the product of the first collapsed fragment might already be
an important source of feedback. Hence, even for the question of  the
efficiency of fragmentation it seems that feedback physics have to be
included. 

Let us assume the first $\sim 200\Ms$ cores fragment to form stars
with 100\% efficiency. If the ratio of produced UV photons per solar
mass is the same as in present day star clusters than about
$6\tento{63}$ UV photons would be liberated during the average life
time of massive star ($\sim 5\tento{7}\yrs$). This is about hundred
times more than the $\sim 4\tento{61}$ hydrogen atoms within the
virial radius.  However, the average recombination time
$(nk_{rec})^{-1}\sim 5\tento{5}\yrs$ within the virial radius is a
factor 100 less than the average lifetime of a massive star. Hence,
very small or zero UV escape fractions for these objects are
plausible. However, the first supernovae and winds from massive stars
will substantially change the subsequent hydrodynamic and chemical
evolution as well as the star formation history of these objects.  A
more detailed understanding of the role of such local feedback will
have to await yet more detailed simulations that include the poorly
understood physics of stellar feedback mechanisms.

Since the collapsing core evolves on much faster timescales than the
rest of the halo it seems plausible that the first star (or star
cluster) will have a mass of less or the order of the core mass. It
seems also quite clear that the radiative feedback from this star
(these stars) will eventually halt further accretion. As a consequence
this might suggest that the formation of very massive objects or
supermassive black holes is unlikely. This later speculation will be
tested by yet higher resolution simulations we are currently working
on.

Recently Bromm, Coppi, and Larson (1999) have studied the
fragmentation of the first objects in the universe. The results of
their simulations using a smooth particle hydrodynamics technique with
isolated boundary conditions disagree with the results presented here.
Their objects collapse to a disk which then fragments quickly to form
many fragments throughout the rotationally supported disk.  The
efficiently fragmenting disk in those simulations originates from the
assumed idealized intial conditions. These authors simulated top-hat
spheres that intially rotate as solid bodies on which smaller density
fluctuations were imposed. Naturally they find a disk. Also it is
clear that for a top-hat if the disk breaks up it will do so
everywhere almost simultaneously. Our results with realistic initial
conditions do not lead to a disk and form the first fragment close to
the center of the halo.

\section{Conclusions}

We have reported first results from an ongoing project that studies
the physics of fragmentation and primordial star formation in a
cosmological context. The results clearly illustrate the advantages
and power of structured adaptive mesh refinement cosmological
hydrodynamic methods to cover a wide range of mass, length and
timescales. All findings of AANZ are confirmed in this study. Among
other things, these are that 1) a significant number fraction of
hydrogen molecules is only formed in virialized halos at the
intersection of filaments, and 2) only a few percent of the halo gas
has cooled to $T\ll T_{vir}$.

The improvement of a factor $\sim 1000$ in resolution over AANZ has
given new insights into the details of the fragmentation process and
constraints on the possible nature of the first structures:
1) Only $\lsim 1\%$ of the baryons within a virialized object can
  participate in population III star formation.
2) The formation of super massive black holes or very massive
  objects in small halos seem very unlikely.
3) Fragmentation via Bonnor--Ebert instability yields a $\sim
200\Ms$ core within one virialized object.
4) If the gas were able to fragment further through 3--body \HH
 association and/or opacity limited fragmentation, only a small
 fraction of all baryons in the universe  will be converted into small
 mass objects.   
5) The escape fraction of UV photons above the Lyman limit should
initially be small due to the high column densities of HI ($N_{HI}\sim
10^{23}\cm^{-2}$) of the parent primordial molecular cloud.
6) The first star in the universe is most likely born close to the
center of its parent halo of $\gsim 10^5\Ms$.

\acknowledgments 

This work is supported in part by NSF grant AST-9803137 under the
auspices of the Grand Challenge Cosmology Consortium (GC$^3$).  NASA
also supported this work through Hubble Fellowship grant
HF-0110401-98A from the Space Telescope Science Institute, which is
operated by the Association of Universities for Research in Astronomy,
Inc under NASA contract NAS5-26555.  Tom Abel acknowledges support
from NASA grant NAG5-3923 and useful discussions with Karsten
Jedamzik, Martin Rees, Zoltan Haiman, and Simon White.

\end{document}